\documentclass[aps,pre,twocolumn,float]{revtex4}
\usepackage{amsmath,bm,epsfig}

\let\*\cdot
\def\<{\left\langle} \def\>{\right\rangle} \def\({\left(} \def\){\right)}
 \let\~\widetilde \let\^\widehat 

\def\be{\begin{equation}}\def\ee{\end{equation}}
\def\bea{\begin{eqnarray}}\def\eea{\end{eqnarray}}
\def\bse{\begin{subequations}}\def\ese{\end{subequations}}
\newcommand{\BE}[1]{\begin{equation}\label{#1}}
\newcommand{\BEA}[1]{\begin{eqnarray}\label{#1}}
\newcommand{\BSE}[1]{\begin{subequations}\label{#1}}

\usepackage{pstricks}
\usepackage{pst-node}
\usepackage[ansinew]{inputenc}
\usepackage{amssymb,amsmath}

\def\BSE{\begin{subequations}}\def\ESE{\end{subequations}}
\let \= \equiv

\def\a{\alpha}

\def\g{\gamma}

\def\o{\omega}
\def\wt{\widetilde}

\def\be{\begin{equation}}       \def\ba{\begin{array}}

\def\ee{\end{equation}}         \def\ea{\end{array}}

\def\bea {\begin{eqnarray}}      \def\eea {\end{eqnarray}}

\def\bean{\begin{eqnarray*}}    \def\eean{\end{eqnarray*}}

\def\e{\varepsilon}           \def\ph{\varphi}

\def\const {\mathop{\rm const}\nolimits}

\def\RA {\ \Rightarrow\ }

\def\<{\langle} \def\({\left(}  \def\>{\rangle} \def\){\right)}

\newtheorem{exi}{Example}

\begin{document}

\title{Energy transport in weakly nonlinear wave systems \\with narrow frequency band excitation}
\author{Elena Kartashova}
 \email{Elena.Kartaschova@jku.at}
  \affiliation{Institute for Analysis, J. Kepler University, Linz, Austria}
   \begin{abstract}
A novel discrete model (D-model)  is presented describing nonlinear wave interactions  in systems with small and moderate nonlinearity under narrow frequency band excitation. It integrates in a single theoretical frame two mechanisms of energy transport between modes, namely intermittency and energy cascade and gives conditions when which regime will take place. Conditions for the formation of a cascade, cascade direction, conditions for cascade termination, etc. are given and depend strongly on the choice of excitation parameters.
The energy spectra of a cascade may be computed yielding discrete and continuous energy spectra. The model does not need statistical assumptions as all effects are derived from the interaction of distinct modes.
In the example given --   surface water waves with dispersion function $\o^2=g\,k$ and small nonlinearity -- D-model predicts asymmetrical growth of side-bands for Benjamin-Feir instability while transition from discrete to continuous energy spectrum excitation parameters properly chosen yields the saturated Phillips'  power spectrum $\sim g^2\o^{-5}$. D-model can be applied to the experimental and theoretical study of numerous  wave systems appearing in hydrodynamics, nonlinear optics, electrodynamics, plasma, convection theory, etc.
\end{abstract}

\maketitle
\tableofcontents

\section{Introduction}

A central topic in the theory of weakly nonlinear wave interactions is the mechanism of energy transport between modes. Considering what we can describe in theory and observe in experiment, there is good reason to believe that in any weakly nonlinear dispersive wave system there are two main types of energy transport: intermittency which is a periodic or chaotic  exchange of energy among a small number of modes, and energy cascade which is a unidirectional flow of energy through scales in Fourier space.

In systems with distributed initial state energy transport is studied in the frame of \emph{kinetic} wave turbulence theory (WTT) by means of the wave kinetic equation, \cite{ZLF92,N11}. In this paper we explicitly study wave systems with \emph{narrow frequency band excitation}.

 \emph{Intermittency} is based on finite size effects in a resonator. The general properties of weakly nonlinear wave systems showing intermittency have first been characterized through the solution of the kinematic resonance conditions, \cite{PRL}, which reflect the geometry of the resonator. The general dynamical characteristics of this type of energy transport have been studied in the frame of \emph{discrete} WTT, \cite{K09b}, for systems with narrow frequency band excitation. Main mathematical object of the discrete WTT is a set of dynamical systems for the amplitudes of interacting waves; each dynamical system corresponds to a resonance cluster composed of a small number of resonant triads or quartets having joint modes, \cite{CUP}.

 \emph{Energy cascades} in systems with narrow frequency band excitation have recently been described in \cite{K12a}  using  increment chain equation method (ICEM). An energy cascade is
 represented as a chain of modes with nonlinear frequencies triggered by modulation instability at each cascade step. The energy spectra $E(\o)$ obtained by the ICEM have \emph{exponential decay} and can be written as:
\be \label{spect-D}
E(\o) \sim \sum_{i=1}^{i \ge 2} C_i \o^{-\g_i}, \ \g_i>0
\ee
 where for given linear dispersion function $\o \sim k^\a$,  $C_i$ are known functions of excitation parameters and   $\g_i$ vary for \emph{different magnitudes of nonlinearity}. For comparison, in the systems with distributed initial state, studied in the frame of \emph{kinetic} wave turbulence theory (WTT), energy spectra decays  according to a power law,
\be \label{spect-KZ}
E(\o) \sim  \o^{-\g}, \, \g>0\ee
with different $\g$ for different wave systems, \cite{ZLF92,N11}.

In this paper we present, based on the resonance conditions,
a common mathematical model, called D-model ("D" for "discrete"), incorporating both forms of energy transport, intermittency and cascades, and give criteria under which conditions to expect which behavior.

In D-model, intermittency occurs for very small nonlinearity, $0< \e < 0.1 $, provided that the geometrical form of the resonator permits resonance. An energy cascade occurs at larger levels  of nonlinearity, $\e \sim 0.1 \div 0.4$, and its spectrum does not depend on  shape or finiteness of the interaction domain. The outcome of the model strongly depends on the excitation parameters.

D-model can explain the following phenomena observed in systems with narrow frequency band excitation:

-- no cascade but recurrent wave patterns are observed, \cite{HH93} (surface water waves);

-- a cascade consisting of two distinct parts -- discrete and continuous; form of spectra does not follow a power law, \cite{Mor08} (thin elastic steel plate); \cite{fauve} (gravity-capillary waves in mercury);

-- a discrete energy cascade develops a strongly nonlinear regime yielding breaking, a continuous part of the spectrum is not observed, e.g. \cite{denis} (surface water waves);

-- form of energy spectra depends on the parameters of excitation, e.g. \cite{LNMD09,XiSP10} (gravity surface and capillary water waves correspondingly);

-- amplitudes of direct and inverse cascades are  not symmetric, e.g. \cite{BF67,TW99,LYRF77,Taiw1,Taiw2};

-- interactions of waves over several orders of magnitude \cite{ABKL09} (capillary waves in helium).

The model is briefly described  in Sec.\ref{s:D-model}.
To demonstrate how D-model works,  we give an example determining cascade direction and scenarios of cascade termination for surface water waves, depending on the excitation parameters (Sec.\ref{s:GravWaves}).

In Sec.\ref{s:Versus} we compare assumptions and predictions of D-model and kinetic WTT to give an experimentalist clues which model to apply in a given experimental set-up.
A short list of conclusions and open questions is given in Sec.\ref{s:SumOpen}.

\section{D-model}\label{s:D-model}
 Time evolution of a wave field in a weakly nonlinear wave system is described by  a weakly nonlinear PDE of the form 
\be \label{Non_eps}L(\psi)= - \e N(\psi)\ee
 where  $N$ is a nonlinear operator,  $0<\e \ll 1$ and $L$ is an arbitrary linear dispersive operator, i.e. $L(\ph)$=0 for Fourier harmonics
$\label{FourHarm} \ph=A \exp{i[ \textbf{k} \textbf{x} -\omega ( \textbf{k}) t]} $
with constant $A$. Here $A,\, \textbf{k}, \, \o=\o( \textbf{k})$ denote amplitude, wavevector and dispersion function correspondingly. The small parameter is usually introduced as  wave steepness $\e=A\,k, \, k=|\textbf{k}|.$ If the nonlinearity is small enough, only resonant interactions have to be taken into account. The resonance conditions read
 \bea
 &\mbox{for 3 waves:\, }&
\begin{cases}\label{res}
\omega ({\bf k}_1) + \omega ({\bf k}_2)= \omega ({\bf k}_{3}), \\
{\bf k}_1 + {\bf k}_2 ={\bf k}_{3}. \label{resn}
\end{cases}\\
 &\mbox{for 4 waves:\, }&
 \begin{cases}\label{res4}
\omega ({\bf k}_1) + \omega ({\bf k}_2)= \omega ({\bf k}_{3}) + \omega ({\bf k}_{4}),\\
{\bf k}_1 + {\bf k}_2 ={\bf k}_{3} +{\bf k}_{4}. \label{resn4}
\end{cases}
\eea
Dynamical systems describing time evolution of the slowly changing amplitudes $A_j$ of resonantly interacting modes can be obtained from (\ref{Non_eps}),(\ref{res}) or (\ref{Non_eps}),(\ref{res4}) using e.g. a multi-scale method. In a 3-wave system $A_j=A_j(T), \, T=t/\e$ and in a 4-wave system $A_j=A_j(\wt{T}), \, \wt{T}=t/\e^2$. The corresponding dynamical systems (in canonic variables) are written out below:
\be
i\dot{A}_1=  Z A_2^*A_3,\,
i\dot{A}_2=  Z A_1^* A_3, \, i\dot{A}_3=  - Z A_1 A_2;\label{ch4:complexB0}\ee
\bea
 \begin{cases}\label{4grav-dynamics}
  i\, \dot{A}_1=  V A_2^*A_3A_4 +(\tilde \omega_1 - \o_1) A_1\,,\\
 i\, \dot{A}_2=V A_1^*A_3A_4 +(\tilde \omega_2 - \o_2) A_2\,, \\
 i\, \dot{A}_3=  V^* A_4^*A_1A_2 +(\tilde \omega_3 - \o_3) A_3\,, \\
 i\, \dot{A}_4=  V^* A_3^*A_1A_2 + (\tilde \omega_4 - \o_4)A_4\,, \\
\tilde \omega_j - \o_j = \sum_{i=1}^4(V_{ij}|A_j|^2 -   \frac12\,  V_{jj}|A_i|^2)\,,
\end{cases}
\eea
where the interaction coefficients $V_{ij}= V_{ji}\= V_{ij}^{ij}$  and $V=V^{12}_{34} $
are  responsible for the nonlinear shifts of frequency and the energy exchange within a quartet correspondingly;
 $ (\tilde \omega_j - \o_j) $
are  Stokes-corrected frequencies. For very small nonlinearity, dynamical system (\ref{4grav-dynamics}) can be regarded in a simplified form, with $\tilde \omega_j - \o_j =0$, i.e. without nonlinear correction of frequencies.

 3-wave interactions  dominate in a weakly nonlinear wave
system if resonance conditions (\ref{res}) have solutions
and the coupling coefficients $Z\neq 0$. Otherwise, the leading nonlinear processes
are 4-wave interactions.

The following results hold likewise for resonances and   quasi-resonances  with small enough frequency mismatch.

\subsection{Intermittency, $0<\e <  0.1 $}\label{ss:interm}
 \emph{Excitation of a single mode}  in a 3-wave system generates energy exchange within a resonance cluster only if this is the high-frequency mode $ \o({\bf k}_{3})$ from (\ref{res}). In a 4-wave system, excitation of a single mode generates energy exchange only if it is the high-frequency mode $ \o({\bf k}_{3})$ in a Phillips quartet
\be
\omega ({\bf k}_1) + \omega ({\bf k}_2)= 2\o ({\bf k}_{3}),\ \
{\bf k}_1 + {\bf k}_2 =2{\bf k}_{3}, \label{Phil-q}
\ee
\noindent
which is a special case of (\ref{res4}),  \cite{Has67}.
Solutions of resonance conditions (\ref{res}),(\ref{res4}) form a set of independent resonance clusters. The form of a cluster uniquely defines its dynamical system.

Solutions of dynamical systems (\ref{ch4:complexB0}),(\ref{4grav-dynamics}) are known, \cite{whitt,SS05};
they describe periodic energy exchange within a resonant triad or quartet correspondingly. Resonance clusters of a more complicated structure may have a dynamical system with periodic or chaotic evolution, \cite{CUP}.

\begin{figure}[hc]
\begin{center}
\includegraphics[width=5cm]{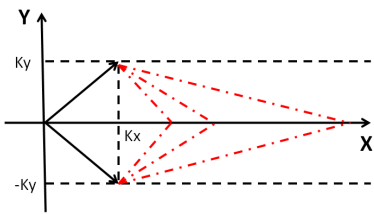}
\caption{\label{f:nonloc} Color online. Nonlocal interactions in a 4-wave system, $\o \sim k^{\a}$. Each couple of (red) dot-dashed lines with equal lengths correspond to specific choice of a parameter $\mathbf{s}$.}
\end{center}
\end{figure}

In both 3- and 4-wave systems,  resonant interactions are not local in $k$-space; even more, in a 4-wave system with dispersion function  $\o \sim k^{\a}$, modes with \emph{arbitrary big difference in wavelengths} can interact directly. In this case a parametric series of solutions of resonance conditions can be easily written out:
\bea \label{nonlocal}
\begin{cases}
k_1^{\a}+k_2^{\a} =k_3^{\a}+k_4^{\a}, \ \
 {\bf k}_1+{\bf k}_2={\bf k}_3+{\bf k}_4, \RA \\
{\bf k}_1= (k_x,k_y), \ \ {\bf k}_2=(\mathbf{s},-k_y), \\ {\bf k}_3=(k_x,-k_y), \ \ {\bf k}_4=(\mathbf{s},k_y),
\end{cases} \eea
where $\mathbf{s}$ is an arbitrary real parameter (see Fig.\ref{f:nonloc}).

In any given 3-wave system, most of the modes are non-resonant.
A non-resonant mode, being excited, does not change its energy at the slow time scale $T$. In the majority of 4-wave systems, each mode satisfies (\ref{res4}). However, excitation of a single mode does not generate resonance in the general case: the excited mode has to be the high-frequency mode in a Phillips quartet.

\subsection{D-cascade, $\e \sim 0.1\div 0.4 $}
 D-cascade  means a cascade computed in D-model by the ICEM method first presented in \cite{K12a}.
In both 3- and 4-wave systems,  D-cascades are generated by  modulation instability (MI). Accordingly, the ICEM method can be applied for all PDEs in which MI has been established: focusing weakly nonlinear Schroedinger equation (NLS), \cite{BF67}; modified NLS, \cite{DY79,Hog85};
modified Korteweg-de Vries equation, \cite{DN76,GPPT01}; and  Gardner equation, \cite{RTP08}.

In both 3- and 4-wave systems,  D-cascades  are generated by MI which is described as a particular case of the Phillips quartet (\ref{Phil-q}) with
 $\o_1=\o_0 + \Delta \o, \, \o_2=\o_0 - \Delta \o, \, 0<\Delta \o \ll 1$:
\be \label{MI}
\o_1 + \o_2 = 2\o_0, \ \
{\bf k}_1+{\bf k}_2=2{\bf k}_0.
  \ee
The mode with  frequency $\o_0$ is called \emph{carrier mode}.  At each step of a discrete cascade, conditions (\ref{MI}) are satisfied, with a \emph{new carrier mode} generated from the previous cascade step.

Time evolution of the quartet (\ref{MI}) is studied in the frame of the nonlinear Schroedinger equation. The corresponding time scale $\tau = t/\e^2$ is  called  Benjamin-Feir time scale  and is shorter than the time scale of resonant interactions. To understand this one has to take into account that small parameter $\e_{res}<0.1$ yielding resonance interactions is in fact substantially smaller than $\e_{MI} \sim 0.1 \div 0.4$ corresponding to modulation instability: $\wt{T}=t/\e_{res}^{2}>t/\e_{MI}^{2}=\tau $. This fact is well established e.g. in the theory of wind generated  oceanic waves, \cite{PJ04}.

Conditions for MI to occur may be given as an instability interval for initial real amplitude $A$ and frequency $\o$ of the carrier wave.
For  the NLS with dispersion relation $\o^2=g\,k$ and small nonlinearity $\e\sim 0.1$ to 0.25  the instability interval is described by
\be
0 < {\Delta \o}/{A k\o} \le \sqrt{2}\label{InstInter}.
\ee
The most unstable mode in this interval  satisfies the so-called maximum increment condition (in Benjamin-Feir form, \cite{BF67}):
\be \label{BFI-incr}
\Delta \o /\o A k  =1.
\ee
 For moderate nonlinearity, $\e\sim 0.25$ to 0.4, the maximum increment condition reads (in Dysthe form, \cite{DY79}),
\be \label{Dys}
\Delta \o/\Big(\o  A k - \frac{3}{2}\o^2  A^2 k^2\Big)=1.
\ee

 Eqs. (\ref{BFI-incr}) and (\ref{Dys}) each generate two chain equations (one for direct D-cascade and one for inverse D-cascade) describing the connection between the amplitudes of two neighboring modes in the D-cascade, under the following assumptions, \cite{K12a}:

(*) the fraction $p$ of energy transported from one cascading mode to the next one depends only on the excitation parameters and not on the step number of the cascade; $p$ is called cascade intensity;

(**) modes forming a D-cascade have maximum instability increment,  i.e. a cascade is formed by the most unstable modes within the corresponding intervals of instability.  This is a mathematical reformulation of the Phillips hypothesis that  the spectral density  is saturated at a level determined by wave breaking, \cite{Ph62}.

In particular, (\ref{BFI-incr}) generates chain equations connecting mode $n$ to mode $n+1$
\bea
 \o_{n+1}=\o_n +  \o_n A(\o_n) k_n, \label{inc-n1-dir}\\
  \o_{n+1}=\o_n -  \o_n A(\o_n) k_n, \label{inc-n1-inv}
\eea
 for direct and inverse D-cascades correspondingly. This means that
the D-cascades are formed by \emph{nonlinear frequencies} depending on the amplitudes.

 From  the chain equations various properties of  D-cascades can be derived, including the form of the discrete and continuous energy spectra.

\section{Surface water waves}\label{s:GravWaves}

 To demonstrate the wide range of the predictions which are given by our model we have chosen a classical example -- surface water waves with dispersion function $\o^2=g\, k$ and small nonlinearity, $\e\sim 0.1\div 0.25$.

Before proceeding with our study we need to make an important remark on the terminology used below. Standard vocabulary for discussing wave resonant interactions
is "a 3-wave system" if (\ref{res}),(\ref{ch4:complexB0}) are satisfied and "a 4-wave system" if (\ref{res4}),(\ref{4grav-dynamics}) are satisfied. Regarding resonance conditions for a Phillips quartet (\ref{Phil-q}) one might formally conclude that this is a system of three waves with frequencies $\o_1$, $\o_2$ and $2\o_3$. However, comparing the dynamical system for a 3-wave system (\ref{ch4:complexB0})  and the dynamical system for a Phillips quartet obtained from (\ref{4grav-dynamics}) by taking $A_3=A_4$ we can see immediately that these systems are different. Accordingly, a Phillips quartet may be referred to in the literature as a 4-wave system.

In the text below we call the system (\ref{MI}) a 4-wave system though in the original papers whose results are interpreted using D-model this system is often called a 3-wave system.
Our terminology also allows us to avoid confusion while discussing cascade termination due to intermittency in Sec.\ref{sss:FPU}.

\subsection{Discrete and continuous energy spectra}\label{ss:spectrum}
For determining D-cascade direction and scenarios of D-cascade termination we need first to compute the form of discrete energy spectrum. Detailed computation of D-spectra for various wave systems are given in \cite{K12a}.
For the convenience of reader below we outline this computation for surface water waves with small nonlinearity.

All computations below are performed with chain equation (\ref{inc-n1-dir}) and yield energy spectra for direct cascade. Computations for inverse cascade should be conducted  similarly but with chain equation (\ref{inc-n1-inv}); they are omitted here.

 Assumptions (*), (**) mean that $E_n=p \,E_{n-1}$ at any cascade step $n$ , $E_n \sim A_n^2$ being the energy of the mode with amplitude $A_n$. As the dispersion function in this case has the form $\o^2=g\, k$,
  this allows to rewrite (\ref{inc-n1-dir}) as
\be \label{dir}
 \sqrt{p} A(\o_n) =  A(\o_n + \o_n^3 A_n/g)= \sum_{s=0} ^ {\infty } \frac {A_n^{(s)}}{s!} \, (\o_n^3 A_n/g)^{s}
\ee
(here notation $A_n=A(\o_n)$ is used).

Restricting ourselves to the first two terms of the Taylor expansion for the left-hand side of (\ref{dir}), we can obtain an ordinary differential equation and solve it analytically:
\bea
\o_n^3 A_n^{'}A_n/g + (1-\sqrt{p})A_n= 0 \RA  \label{2terms-0}\\
 A_n = g\,\frac{(1-\sqrt{p})}{2} \o_n^{-2} + C,\,
 C= A_0 -g\, \frac{(1-\sqrt{p})}{2} \o_0^{-2}.  \label{C-dir}
\eea
Accordingly,  \emph{the discrete energy spectrum} for the direct cascade reads
\be
E_n=E(\o_n) \sim A_n^2= g^2\,\Big[ \frac{(1-\sqrt{p})}{2} \o_n^{-2}+C \Big]^2,\label{energy-grav-Direct-om}
\ee
 where $\o_0, A_0$ are the excitation parameters and $p=p(\o_0, A_0)$.

 The corresponding \emph{continuous energy spectrum} $E(\o)$ is computed as  $\lim_{n\rightarrow \infty}|E_{n+1}-E_n|/|\o_{n+1}-\o_n|$ yielding
\be
E(\o)
\sim 2 g^2\,\Big[ (1-\sqrt{p}) \o^{-5} - C\o^{-3} \Big]. \label{cont-small-gr}
\ee

In particular, the special choice of excitation parameters $C=0$ yields
\be
E(\o) \sim g^2  \o^{-5} \label{cont-Phil}
\ee
which  is the saturated  Phillips' spectrum, \cite{Ph62};  this is also in accordance with the JONSWAP spectrum (an empirical relationship based on experimental oceanic data).

Kinetic WTT predicts $\sim \o^{-4}$ in this case,  \cite{ZLF92,N11}.

\subsection{Cascade direction}
Combining chain equation and  expression for the amplitudes of the cascading modes  we can study how cascade direction depends on the choice of excitation parameters.

For instance, for direct cascade $\o_{n+1}-\o_n>0$ with $C\neq 0$ the use of (\ref{inc-n1-dir}),(\ref{2terms-0}),(\ref{C-dir}) yields
\bea
0< \o_{n+1}-\o_n=\o_n^3 A(\o_n)/g=  \\
\o_n^3 \Big[ g\,\frac{(1-\sqrt{p})}{2} \o_n^{-2}+C \Big]/g = \label{cond-dir-cas-zero} \\
\frac{(1-\sqrt{p})}{2} \o_n +  \Big[ A_0 - g\,\frac{(1-\sqrt{p})}{2} \o_0^{-2}\Big]\o_n^3/g =  \\
\frac{(1-\sqrt{p})}{2} + \Big[ A_0 - g\,\frac{(1-\sqrt{p})}{2} \o_0^{-2}\Big]\o_n^2/g \RA  \\
 A_0 - g\, \frac{(1-\sqrt{p})}{2} \o_0^{-2} > 0  \RA \\
g\, (1-\sqrt{p}) +\Big[ 2 A_0 - g\,(1-\sqrt{p}) \o_0^{-2}\Big]\o_n^2 >0 \label{cond-dir-cas}
\eea
As $(1-\sqrt{p})>0,$ the range of frequencies forming direct cascade depends only on the sign of the expression
$ 2 A_0 - g\,(1-\sqrt{p}) \o_0^{-2}$.

An easy examination of (\ref{cond-dir-cas-zero}),(\ref{cond-dir-cas}) shows how to choose excitation parameters $A_0, \, \o_0$ in order to observe direct cascade:
\be \mbox{if  } \quad  2 A_0 \ge g\,(1-\sqrt{p}) \o_0^{-2},\ee
the only restriction on the range of frequencies forming direct cascade is trivial:  $\o_n >\o_0$; accordingly, only direct cascade will occur;
\be \mbox{if  } \quad  2 A_0 < g\,(1-\sqrt{p}) \o_0^{-2},\ee
direct cascade will be observed for the  range of frequencies $\o_0< \o_n \le \o_{n_{st}}$ where
\be
\o_{n_{st}}=\sqrt{\frac{g\,(1-\sqrt{p})}{g\,(1-\sqrt{p})\o_0^{-2}-2A_0}}.\label{dir-range-all}
\ee
For simplifying further formulae we introduce here a small parameter $\e_0=A_0k_0=A_0\o_0^2/g$ and rewrite (\ref{dir-range-all}) as
\be
\o_{n_{st}}=\o_0\sqrt{\frac{(1-\sqrt{p})}{(1-\sqrt{p})-2\e_0}}.\label{dir-range-eps}
\ee
Physical meaning of the frequency $\o_{n_{st}}$ is explained in  Sec.\ref{ss:stabil}.

Similar computations can be performed for inverse cascade, and also the case when both direct and inverse cascade are possible can be studied this way. In particular, for some choice of excitation parameters both direct and inverse cascade can be initiated simultaneously.
This scenario is supported by wide range of experimental studies, e.g. \cite{TW99,Taiw1,Taiw2}.

All formulae (\ref{2terms-0}),(\ref{C-dir}),(\ref{dir-range-eps}) are given in terms of excitation parameters $A_0, \, \o_0$ and cascade intensity $p$. This means that we should also compute $p$ as a function of $A_0, \, \o_0$, $p=p(A_o,\o_0)$. This tedious computation will be given elsewhere. However, in the next section we give an example of the computation for a particular form of the solution (\ref{2terms-0}).

Notice that for studying predictions of the D-model in experimental data one can just measure $\sqrt{p}$ as the ratio of amplitudes of two consecutive cascading modes, $\sqrt{p}=A_{n+1}/A_n$, and apply formulae afterwards.

\subsection{Cascade termination}
\subsubsection{Breaking}\label{ss:growth}
It was first shown in \cite{DP86} that the amplitude of the carrier wave may become so large that its steepness exceeds locally the maximum
steepness of gravity waves yielding  the onset of wave breaking.

In order to demonstrate that this effect can be reproduced in D-model,
let us regard a particular solution of  (\ref{2terms-0}) with $C=0$:
\be
A_n = g\,\frac{(1-\sqrt{p})}{2} \o_n^{-2}\label{gr00}.
\ee
As for this solution
\bea
A_0 =g\, \frac{(1-\sqrt{p})}{2} \o_0^{-2} \RA  \label{gr0} \\
\begin{cases}\label{gr1}
p=(1-2\e_0)^2 \,   \\
A_n=p^{n/2}A_0 =(1-2\e_0)^{n}A_0,
\end{cases}
\eea
any choice of $\e_0$ and $A_0$ defines uniquely a cascade intensity $p$ and the amplitude of the $n$-th cascading mode.

It follows from (\ref{gr00}),(\ref{gr0}) that in this case all cascading modes have the same steepness $\e_n=\e_0, \, \forall n$:
\be
\e_n=A_nk_n=A_n\o_n^{2}/g=\frac{(1-\sqrt{p})}{2}=\e_0.
\ee
This allows to compute the steepness $\e$ of the total wave packet at  step $n$ (before breaking) as
\be
\e \approx \sum_{n}\e_n \approx (n+1)\e_0.
\ee
Accordingly, though the amplitudes of the cascading modes are decreasing, \emph{the steepness of the total packet is growing} with an increasing  number of cascade steps.

For instance, direct computations demonstrate that if initial steepness $\e_0=0.1$, then after 3 cascade steps $A_0 \cdot 100\%/A_3 \approx 0.5\%$. However, the total steepness of the wave packet is $\e = 4\cdot 0.1 \sim 0.4$ and according to the Stokes criterion for the limiting steepness being about 0.44, we conclude that  mode $A_3$ is about to break. A different choice of the initial steepness, say $\e_0=0.05$, yields the same total steepness $\e =8 \cdot 0.05 \sim 0.4$ at the step $n=7$ and  cascading mode $A_7$ contains about $23\%$ of the excitation energy while $A_0 \cdot 100\%/A_7 \approx 48\%$. Thus, varying excitation parameters one can predict the occurrence of breaking  at the different cascade steps.

Denoting  limiting steepness of the wave package before breaking as $\e_{br}$, we conclude that the cascade terminates
due to breaking if
$
(n_{br}+1)\e_0 = \e_{br} \approx 0.44,
$
i.e. at \emph{the finite step} $n_{br}$,
\be
n_{br} \approx 0.44/\e_0 -1. \label{stab-cond-Phil}
\ee

At the end of this section we point out again that all results given by (\ref{gr0})-(\ref{stab-cond-Phil}) are obtained for a specific form of solution of (\ref{2terms-0}), namely, for $C=0$.

In the general case $C \neq 0$ some results might be qualitatively different: for instance, breaking may occur in the infinity rather than at some finite step.

In this section we did not aim to present all possible formulae in their most general form but rather to demonstrate that growth of nonlinearity following by breaking  -- an experimentally well established phenomenon, \cite{denis,TW99,Taiw1,Taiw2} --
can be reproduced by the D-model.

\subsubsection{Stabilization}\label{ss:stabil}
If at some cascade step $n_{st}$ the mode with frequency $\o_{n_{st}}$ is stable, then
the condition (\ref{InstInter}) is not fulfilled, no additional mode can be generated and the D-cascade stops due to stabilization at some frequency $\o_{n_{st}}$.

From (\ref{InstInter}),(\ref{2terms-0}),(\ref{C-dir}) it may be concluded that
\bea
\o_{n_{st}}=\o_{n_{st}+1} \RA 0=\o_{n_{st}}-\o_{n_{st}+1}= \\
A_{n_{st}}\o_{n_{st}} k_{n_{st}}=\Big[ g\,\frac{(1-\sqrt{p})}{2} \o_{n_{st}}^{-2}+C \Big]\o_{n_{st}}^3/g \RA \\
0 = \frac{(1-\sqrt{p})}{2} \o_{n_{st}} + C\o_{n_{st}}^3/g \RA \eea
\bea \o_{n_{st}}^2=\frac{(1-\sqrt{p})}{2} /C = \frac{(g\,1-\sqrt{p})}{g\,(1-\sqrt{p})\o_0^{-2}-2A_0}.
\eea
and for direct cascade  stabilization occurs if
\be
\o_{n}>\o_{n_{st}}=\o_0\sqrt{\frac{(1-\sqrt{p})}{(1-\sqrt{p})-2\e_0}},\label{cas-term-2-dir}
\ee
which is in accordance with (\ref{dir-range-eps}).

It follows from (\ref{cas-term-2-dir}) that direct cascade

\textbf{(a)} stabilizes \emph{at the finite step} $\o \le \o_{n_{st}}<\infty$ if  $1-\sqrt{p}>2\e_0;$

\textbf{(b)} stabilizes \emph{in infinity} if $1-\sqrt{p}=2\e_0;$ then $C=0$ in (\ref{C-dir}) and
corresponding continuous energy spectrum is Phillips spectrum $\sim \o^{-5}$ (see Sec. \ref{ss:growth});

\textbf{(c)} stabilization does not occur if  $1-\sqrt{p}<2\e_0$  while expression on the RHS of (\ref{cas-term-2-dir}) becomes complex and has no physical meaning, i.e. stabilization conditions can never be fulfilled.

Similar computations can be performed for inverse cascade. Though formally the termination conditions may allow the inverse cascade to be terminated at a negative frequency,  this is physically irrelevant. This means that in a real physical system an inverse  cascade terminates in some vicinity of zero frequency mode which might yield a substantial  concentration of energy near zero frequency mode, also observed experimentally, e.g. \cite{XiSP10}.

\subsubsection{FPU-like recurrence}\label{sss:FPU}
The fact that the long-time evolution of nonlinear wave trains of surface water waves may evolve in recurrent fashion (FPU-like recurrence), where the wave form returns periodically to its previous form, has been discovered experimentally and described in the pioneering paper of Lake et al., \cite{LYRF77}. The next mile-stone step in the study of this effect has been performed by Tulin and Waseda in \cite{TW99} where the authors refined the experimental technique in a way that not only excitation frequency but also initial side bands and the strength of amplitude could be chosen. More experimental results can be found in \cite{Taiw1,Taiw2} and bibl. therein.

In  D-model, formation of a recurrent phenomenon (intermittency) is due to  formation of a  cluster of resonant quartets, in the simplest case -- an isolated Phillips quartet, (\ref{Phil-q}). Its occurrence depends strongly on the form of the experimental tank.

 For some aspect ratio of the tank side lengths, intermittency can not occur as kinematic resonance conditions can not be satisfied. If for given aspect ratio, solutions of (\ref{res4}) exist, interaction coefficient $V \neq 0$ and initially excited resonant mode(s) are modulationally  stable, then a recurrence may be observed.

Below we give a short list of experimental observations with their respective explanations:

-- \emph{no cascade is observed}, rather recurrent patterns on the water surface are observed, \cite{HH93}:

 initial steepness is too small to initiate modulation instability;

-- \emph{no intermittency is observed, rather a discrete cascade terminated by wave breaking}, \cite{denis}:

 initial steepness is big enough to cause modulation instability and $\o_{br} < \o_{st}$ or stabilization is generally not possible for the chosen excitation parameters;

-- \emph{no intermittency is observed in the non-breaking regime}, \cite{Mel82}:

 initial steepness is big enough to cause modulation instability, cascade terminates due to stabilization, i.e. $\o_{st} < \o_{br}$, and the mode with frequency $\o_{st}$ is not a resonant mode in a resonant cluster possible for chosen experimental tank;

-- \emph{intermittency is observed in the non-breaking regime}, \cite{Taiw1,Taiw2}:

cascade stabilizes at the frequency $\o_{st}$, the $\o_{st}$-mode is resonant mode and may excite a resonant cluster with another cascading mode. In particular, if $\o_{st}$-mode and $\o_0$-mode form a resonance, complete FPU-like recurrence will be observed, \cite{TW99,Taiw1,Taiw2}.
If $\o_{st}$-mode forms a resonance with cascading mode with frequency $\wt{\o} \neq \o_0$, then partial recurrence will occur, with spectral peak being downshifted to the frequency $\wt{\o}$, \cite{Mel82}.

-- \emph{intermittency is observed at post-breaking stage}, \cite{TW99,Taiw1,Taiw2}:

as essential part of the energy is lost due to  breaking, amplitudes of newly excited modes may become modulationally stable and form a resonance with some of the previously excited cascading modes. This is only a qualitative explanation, quantified prediction is an important separate topic which lies outside the scope of this paper. A possible theoretical scenario of the energy redistribution at the post-breaking stage is developed in \cite{TW99}.

In this section we have shown how to use the chain equation to determine, depending on the excitation parameters, the direction of the energy cascade and how the cascade will terminate.

It should be noted that also the asymmetry of direct and inverse cascades as known from experiment, e.g. \cite{TW99,LYRF77,Taiw1,Taiw2}, may be deduced from the chain equation, \cite{KSh11}.

  \begin{widetext}
\begin{table*}
\begin{tabular}{|l|l|l|l|}
\hline
 &property& D-model & kinetic WTT\\
\hline\hline
assumptions & & &\\
\hline
1 &cascade origin& modulation instability, & \emph{\textbf{S}}-wave kin. eq.,\\
  & in an         \emph{\textbf{S}}-wave system       & no dependence on \emph{\textbf{S}} & depends on \emph{\textbf{S}}\\
    &            &  & \\
\hline
2 &initial state&narrow frequency band & distributed state \\
\hline
3 &locality of interactions&  no assumptions & necessary \\
\hline
4 & existence of &  & \\
 & inertial interval& no assumptions & necessary \\
 & &  &  \\
\hline
5 & origin of &  & \\
 &cascade termination& no assumptions & dissipation \\
 & &  &  \\
 \hline
6 & range of waves &  & \\
 &steepness& $0<\e \sim 0.1 \div 0.4 $& $0<\e < 0.1$  \\
 & &  &  \\
  \hline
7 & cascade intensity & is constant & no assumptions\\
 & &  &  \\
  \hline
8 &energy flux& no assumptions& is constant \\
 & &  &  \\
\hline \hline
predictions & & &\\
\hline
1 &cascade is formed  by &  nonlinear frequencies&  linear frequencies\\
\hline
2 &spectrum form & &\\
 &(a)& discrete and continuous,& continuous,\\
 &(b)& depends on   & does not depend  \\
  && the excitation  & on the excitation \\
\hline
3&transition from  &  &  \\
&discrete to &  &  \\
&continuous spectrum& included & not included \\
\hline
4& direction of cascade & included & included\\
\hline
5 &intermittency& included & not included\\
\hline
 6 & origin of &  & \\
   & cascade termination & various scenarios: & (see assumptions)\\
      &  & stabilization, & \\
         &  & breaking, & \\
            &  & FPU-like recurrence & \\
\hline
\end{tabular}
\caption{ \label{t:properties} Assumptions and predictions used in  D-model and  kinetic WTT}
\end{table*}
\end{widetext}

\section{D-model \emph{versus} kinetic WTT}\label{s:Versus}

During almost fifty years, kinetic WTT which requires a distributed initial state, was used to describe experiments using narrow frequency band excitation. This was considered legal, as the assumption was and still is that from the excitation frequency as a starting point quickly a distributed state will establish. The discrete part of the spectrum which was well observed in experiment was ignored in theoretical discussion, focusing on the continuous part of the spectrum.

That this approach is not without problems was acknowledged within the community. As A. Newell noticed recently,  "numerics seems to agree with the theory but experiments not", \cite{N12_France} (see also recent review \cite{NR11}). Indeed, a distributed initial state as needed for applicability of kinetic WTT is easy to create in  numerical simulations but not in laboratory experiments.

Though D-model and kinetic WTT differ greatly in their assumptions and consequent range of applicability, sometimes the predicted form of continuous energy spectrum is very close. To get more understanding which approach to apply in a given experimental setup we give the following comparison of  the assumptions and predictions of  D-model and kinetic WTT (short list is given in  Table \ref{t:properties}).

The crucial difference between  descriptions of energy cascades in  D-model  and in  kinetic WTT is \emph{the physical mechanism generating a cascade}: modulation instability in arbitrary $s$-wave system \emph{versus} $s$-wave interactions,  $s=3, 4,...$ .

This means in particular that a D-cascade is generated by a mechanism which provides locality of interactions automatically. In the kinetic WTT the locality has to be assumed, and no mechanism is suggested which allows to choose local interactions in wave systems where also nonlocal interactions are possible, as  was shown in  Sec.\ref{ss:interm}, Eqs.(\ref{nonlocal}), and is also experimentally observed, \cite{ABKL09}. The assumption of locality -- only interactions among waves with close wavelengths are allowed -- is  basic in the kinetic WTT;  without locality energy exchange among different scales $k$ is possible and the energy spectrum  can not be regarded  as a function of only $k$.

Another important point is that the influence of the excitation parameters on the form of the continuous energy spectrum, observed  experimentally, e.g. \cite{XiSP10,ParEx1,ParEx2,ParEx3}, principally can not be included into kinetic WTT but is reproduced in D-model.

One more considerable difference between D-model and kinetic WTT is the origin of cascade termination. In kinetic WTT this is always dissipation while in D-model various scenarios can be reproduced depending on the excitation parameters and direction of the cascade. D-cascades can terminate e.g.  due to breaking, stabilization or formation of the Fermi-Pasta-Ulam-like recurrent phenomenon; all these effects are observed experimentally, \cite{TW99,Taiw1,Taiw2}.

Assumption (*) of D-model about constant cascade intensity, $p=\const$, is absent in the kinetic WTT. This assumption is not substantial for D-model and can easily be removed. Indeed, if cascade intensity at step $n$ is $p_n \neq \const$,  chain equations  (\ref{inc-n1-dir}),(\ref{inc-n1-inv})
do not change, while the ODE (\ref{2terms-0}) and its solutions can be trivially rewritten by the changing $p$ to $p_n$. The only non-trivial change would be the construction of the transition from discrete to continuous energy spectra. Of course, the estimates  for determining cascade direction, termination, etc. should be recalculated and might get a more complicated form though not necessarily. For instance, all estimates made for the particular solution of (\ref{2terms-0}) with $C=0$ remain valid while for so chosen excitation parameters $A_0, \o_0$
cascade intensity is a constant defined by $A_0, \o_0$:
\bea
C=0 \RA A_0 -g\, \frac{(1-\sqrt{p_n})}{2} \o_0^{-2}=0 \RA \\
p_n=\sqrt{1-2A_0\o_0^2/g} \equiv \const. \label{Phil-p}
\eea
Accordingly, transition from discrete to continuous spectrum can be performed as above producing saturated Phillips spectrum.

A wide range of experimental data shows that $p=\const$ in various wave systems and accordingly the discrete energy spectrum has exponential form, e.g. \cite{ShXP12} and bibl. therein; this was our motivation for choosing constant cascade intensity in this presentation.

Last but not the least. As it was shown in a recent experimental study of capillary waves, "from the measured wavenumber-frequency spectrum it appears that the [linear] dispersion relation is only satisfied approximately. (...)  This disagrees with weak wave turbulence theory where exact satisfaction of the dispersion relation is pivotal.  We find approximate algebraic frequency and wavenumber spectra but with exponents that are different from those predicted by weak wave turbulence theory", \cite{SWW-09}.

 On the other hand,
 D-cascades are formed by the modes with nonlinear frequencies and not by the modes with linear frequencies as it is assumed in kinetic WTT.

This is a manifestation of the very important difference between cascades in the D-model and kinetic WTT. Cascades in the kinetic WTT are due to resonant interactions and therefore are possible at the time scales $T$ or $\tilde{T}$ with very small nonlinearity $0< \e < 0.1.$  In D-model only intermittency is formed at these time scales while D-cascade occurs at the faster time scale $\tau$ and for bigger nonlinearity
  $\e \sim 0.1 \div 0.4$.

\section{Conclusions and open questions}\label{s:SumOpen}

In this paper we presented D-model which describes nonlinear wave systems with narrow frequency band excitation. It allows to reproduce in a single theoretical frame various nonlinear wave phenomena, in particular finite-size effects in resonators and formation of energy cascades. The cascades do not depend on shape or finiteness of the interaction domain as they are triggered by the  local mechanism of modulation instability.

The main predictions of D-model can be stated as follows:

-- Intermittency is formed by a set of distinct modes with \emph{linear frequencies}; intermittency may occur in systems with very small nonlinearity, $0<\e < 0.1$, at the slow time scales $T$ or $\tilde{T}$; the underlying physical mechanism is resonant wave interaction.

-- An energy cascade is formed by a chain of distinct modes with  \emph{nonlinear frequencies}; a cascade may occur in systems with small to moderate nonlinearity, $\e \sim 0.1 \div 0.4$,
at the Benjamin-Feir time scale $\tau$; the underlying physical mechanism is modulation instability.

-- The discrete and continuous energy spectra of a cascade  can be computed by the increment chain equation method, \cite{K12a}; the form of spectra, cascade direction and scenario of cascade termination depend on the excitation parameters.

-- Various scenarios of energy cascade termination, known from laboratory experiments -- stabilization,  breaking and appearance of Fermi-Pasta-Ulam-like recurrence, can be reproduced in D-model.

As it was discussed in Sec.\ref{s:Versus}, all these predictions are quite different from those of kinetic WTT developed for wave systems with distributed initial state. In the latter case an
 energy cascade occurs at the slow time scale of resonant interactions, is formed by linear frequencies, terminates (by assumption) always due to dissipation, etc.

D-model explains known physical phenomena, as well as the results of individual laboratory experiments.
In addition D-model makes  predictions which may be easily verified in experiment, e.g. increasing the amplitude of excitation increases the distance between cascading modes in $k$-space (direct consequence of the chain equation), and others.

It should be mentioned that within the wide range of excitation-dependent spectra predicted by D-model, the saturated Phillips spectrum $\o^{-5}$ has two special properties. Firstly, as  shown in Sec.\ref{ss:stabil}, of  all possible spectra  only the Phillips spectrum  does not stabilize after a finite number of cascade steps but in infinity (in $k$-space). Secondly, for the Phillips spectrum it is easy to prove that cascade intensity is constant (see (\ref{Phil-p})); for other spectra it is not known. What this physically means is presently under study.

D-model may be refined in many ways, e.g.

 -- in (\ref{2terms-0}) just two terms of the Taylor expansion  are taken to compute the energy spectrum; instead, one may regard the hierarchy of finite-order ODEs obtained by cutting off the Taylor expansion at 3, 4 and so on terms;

-- dissipation (depending on frequency) can be taken into account in the following way: cascade intensity $p$ which describes the fraction of energy going from mode $n$ to mode $n+1$ may be considered as a function increasing with frequency, $p=p(\o)\neq \mathbf{const}$. So stabilization of the  cascade will occur earlier and also the form of the energy spectrum will change.

Many more problems can be studied in the frame of D-model than have been mentioned in this paper. For instance,

-- Is it possible to use D-model for describing real-life phenomena where excitation parameters are not \emph{a priori} known?

Most naturally one might study the probability  of various initial states in a given situation and choose as input for the model either the most probable state or an average state -- for instance the known prevailing  direction of the wind blowing over the ocean during a season.

-- Modulation instability plays a central role in the formation of extreme waves, e.g. \cite{freak-theor3,MO-all09,MO-all09-1}. Is it possible to use D-model to predict freak waves in the ocean?

The Benjamin-Feir index (BFI), which is a ratio of the parameter of nonlinearity $\e$ to the
relative spectral width, characterizes the evolution of an unidirectional wave field
with a  narrow spectrum. As either the frequency range or the directional
spreading widen, the probability of appearance of extremely steep waves decreases, \cite{MO-all09,MO-all09-1}. Using chain equation, one may e.g. to compute an upper estimate for BFI at each cascade step as a function of excitation parameters and to study characteristic behavior of this function.

-- In the special case of surface water waves the Zakharov equation is the model of choice. So it would be of great interest to compare the predictions of D-model with predictions of the Zakharov equation.

Some results are known already, for instance, sideband asymmetry of Benjamin-Feir instability is  established in numerical simulation with the Zakharov equation, \cite{SS87}. Moreover,
it was recently shown by M. Onorato, \cite{Onorato12}, that a D-cascade has a direct correspondence in the Zakharov equation: the frequencies of cascading modes as determined in D-model form exact 4-wave resonances in the Zakharov equation  with nonlinear Stokes corrected frequencies.

This result is of the upmost importance as it opens a broad avenue for further studies of nonlinear wave systems with higher degree of nonlinearity. The question is:

\emph{Is it possible  to compute energy cascades in nonlinear wave systems with distributed initial state using  a new type of wave kinetic equation based on resonances of nonlinear Stokes corrected frequencies, with bigger nonlinearity than is possible for applicability of kinetic WTT?}

{\textbf{Acknowledgements.}} Author acknowledges  K. Dysthe, A. Maurel, A. Newell,
M. Onorato, E. Pelinovsky, I. Procaccia, M. Shats, I. Shugan and H. Tobisch  for valuable discussions and anonymous  Referees for useful remarks and recommendations. This research has been supported by
the Austrian Science Foundation (FWF) under project
P22943-N18, and in part -- by the Project of Knowledge Innovation Program (PKIP) of Chinese Academy of Sciences, Grant No. KJCX2.YW.W10.

\end{document}